\documentclass[12pt]{article}
\usepackage{amsthm,amsmath,amsfonts,amssymb,cases}
\usepackage{amsmath,amsfonts,amsthm}
\newcommand{\R}{{\mathord{\mathbb R}}}

\def\lam {\lambda}
\def\g   {\gamma}
\def\a   {\alpha}
\def\b   {\beta}

\newcommand{\HH}{\mathcal{H}}
\newcommand{\DD}{\mathcal{D}}
\newcommand{\FF}{\mathcal{F}}

\newcommand{\WW}{\mathcal{W}}
\newcommand{\RR}{\mathcal{R}}

\def\i {{i}}
\def\e {{e}}
\def\d {{d}}
\newcommand{\ran}{{\rm Ran}}

\newcommand{\w}[1]{\widehat{#1}}
\newcommand{\ov}[1]{\widetilde{#1}}

\def\one{{\sf 1}\mkern-5.0mu{\rm I}}

\newcommand{\ben}{\begin{displaymath}}
\newcommand{\een}{\end{displaymath}}
\newcommand{\beqn}{\begin{equation}}
\newcommand{\eeqn}{\end{equation}}

\def\inf{{\rm inf}\,}

\def\const{{\rm const}\,}
\def\dist{{\rm dist}\,}
\def\supp{\operatorname{supp}}

\newcommand{\der}[1]{\frac{\partial}{\partial #1}}
\newcommand{\sfrac}[2]{\textrm{\footnotesize $\frac{#1}{#2}$}}

\newtheorem{lemma}{Lemma}
\newtheorem{theorem}[lemma]{Theorem}

\newtheorem{proposition}[lemma]{Proposition}

\title{Existence of the $D0$--$D4$ Bound State: \\
a detailed Proof\,\footnote{Work 
partially supported by NSF grant DMS-0200235, by EU
grant HPRN-CT-2002-00277, by \mbox{MaPhySto} -- A Network in Mathematical
Physics and Stochastics, funded by The Danish National Research
Foundation, and by grants from the Danish research council. }}

\author{\hspace{-.2 cm} L. 
Erd\"os${}^{a } $
\footnote{On leave from School of Mathematics, Georgia Tech.} \, , 
D. Hasler${}^{b}$ , \ 
J.P. Solovej${}^{c}$\\ 
\normalsize\it \hspace{-.5 cm}\\
\hspace{-.5 cm}\normalsize\it ${}^{a}$ 
Mathematisches Institut, University Munich \\\normalsize\it
Theresienstr. 39, D-80333, Munich Germany
\\
\normalsize\it \hspace{-.5 cm}${}^{b}$ 
Department of Mathematics, University of British Columbia
\\ \normalsize\it
V6T 1Z2, Vancouver, BC, Canada, 
\\
\normalsize\it \hspace{-.5 cm}${}^{c}$ 
Department of
Mathematics, University of Copenhagen \\  \normalsize\it
Universitetsparken 5, DK-2100 Copenhagen, Denmark}

\date{}

\begin{document}
\maketitle

\begin{abstract}
We consider the supersymmetric quantum mechanical system which is obtained
by dimensionally reducing $d$=6, $N$=1 supersymmetric gauge theory with
gauge group $\mathrm{U}(1)$ and a single charged hypermultiplet.
Using the deformation method and ideas introduced by Porrati and
Rozenberg \cite{porroz:bou}, we present a detailed proof of the 
existence of a 
normalizable ground state for this system.
\end{abstract}

\section{Introduction}\label{sec:int}

The particular system, which we will consider, belongs to a class of 
supersymmetric quantum mechanical models.
These models appear in the study 
of quantized membranes \cite{c4a}, $D$-brane bound states \cite{c5a}, 
and M-theory \cite{c5b}.
Especially the question of existence respectively absence of 
normalizable ground states, i.e., zero energy states, is
of physical importance.
The 
Hamiltonian of these models is of the form
\ben
H = - \Delta + V + H_F \; .
\een
The scalar potential $V$ is polynomial in the bosonic degrees
of freedom and admits zero energy valleys extending to 
infinity while $H_F$ is quadratic in the fermionic degrees
of freedom and linear in the bosonic degrees of freedom. 
Moreover, the Hilbert space
carries a unitary representation of a gauge group.
The physical Hilbert space consists of gauge invariant states.
Due to supersymmetric cancellations, 
the zero energy valleys render the Hamiltonian to have continuous
spectrum, which covers the positive real axis. Therefore, the 
Hamiltonian is non-Fredholm and the question about existence
of ground states is subtle.
The Witten index $I_W$, i.e., the number of bosonic ground states minus
the number of fermionic ground states, can be calculated by means of
\ben
I_W = \lim_{R \to \infty} \lim_{\beta \to \infty} 
\mathrm{Tr} ( (-1)^F \chi_R \mathrm{e}^{-\beta H}) \; ,
\een
where $\chi_R$ denotes the characteristic function of the 
ball of radius $R$ centered around the origin in configuration
space, c.p. \cite{setste:aco}. Since there
is no gap in the spectrum one has to deal with a delicate
analysis of boundary contributions. 
As a different approach, Porrati and Rozenberg proposed in
\cite{porroz:bou}
a deformation 
method to detect the 
existence of normalizable ground states for systems with at
least two real supercharges. 
One deforms the supercharges of the system with a real potential $w$,
\ben
D \rightarrow D_w := e^{-w} D e^{w}  \; , \quad D^{\dagger} \rightarrow
D_w^{\dagger} := e^{w} D^{\dagger} e^{-w}  \; ,
\een
such that the spectrum of the deformed Hamiltonian
$H_w := D_w D_w^{\dagger} + D_w^{\dagger} D_w$
becomes discrete.
This
might allow one to show the existence of a ground state $\Psi_w$ 
for the deformed problem. Provided that $e^{\pm w} \Psi_w$ is 
normalizable, then, the original problem admits a ground state as well.
Using this method  the number of ground states
for numerous models could be determined, \cite{kacsmi:nor}.

In this paper, we consider the quantum mechanical system which is obtained by
dimensionally reducing $N=1$ supersymmetric
gauge theory, with gauge group
$\mathrm{U}(1)$ and with a single charged hypermultiplet
from six dimensions.
The system appears in the problem of counting H-monopole ground states
in the toroidally compactified heterotic string \cite{wit:sma}.
Moreover, the same system describes the low energy dynamics of a 
$D0$--brane in the presence of a $D4$--brane 
\cite{berdou:fiv,doukabpoushe:dbr}.
String duality arguments predict the existence of exactly one bound state at 
threshold for this system, c.p. \cite{pol:supII}. 
The existence of such a state provides a check of 
the correctness of these duality hypotheses.

In \cite{setste:aco}, an analysis was sketched of how to obtain
the value one for the Witten index for this system.
Combined with vanishing Theorems,
\cite{setste:inv}, 
such a result implies that the model has a unique ground
state. 
Independently of the work in \cite{setste:aco}, it was argued
in \cite{porroz:bou} how a deformation method may be used to establish
existence of a ground state. 
In this paper we use this deformation method and follow the main ideas 
of  \cite{porroz:bou} to present a rigorous proof of the existence of a 
ground state.
In particular, we make the argument in \cite{porroz:bou} mathematically 
precise in two important aspects. 
First we prove the existence of a ground state for the deformed
problem: we have to do semiclassical analysis on the space of gauge invariant
functions and we have to deal with the fact that $H_F$ is unbounded.
In a second part we 
prove a decay estimate for the ground state of the 
deformed problem. In particular, we show that it decays
sufficiently fast implying that the original problem also
has a ground state. 
To obtain this decay property, we use an Agmon  \cite{agmon:lec}
estimate and combine it with a symmetry argument.
We think this is a clear and direct way to obtain the  necessary 
decay.
Alternatively one could also determine the asymptotic
form of the ground state by analysing the effective
dynamics along a potential valley. Such an analysis
was indicated in \cite{porroz:bou}. Similarly
one could use a supercharge
analysis related to the one in  \cite{fghhy} (which was
used to determine the asymptotic form 
of the bound state of two $D0$-branes).
Similar considerations have to be taken into account
when using the deformation method to study the number of zero energy states
for other supersymmetric models of the same type.
Moreover, there are results about the structure of the $D0$-$D4$ bound state 
\cite{setste:d0d}.

The paper is organized as follows. In Section 2, we describe the 
model. In Section 3, we introduce the deformation method
and give an outline of the proof, which is then presented in Section 4.

\section{The model} \label{sec:the}

The model is
obtained by dimensionally reducing $N=1$, ${\rm U}(1)$ supersymmetric
gauge 
theory with a single charged hypermultiplet,
from $5+1$ dimensions to $0+1$ dimension
\cite{berdou:fiv,setste:d0d}. The
bosonic
degrees of freedom are given by
\ben
q = (q_j)_{j=1,...,4} \in \R^4 \; , \ \ \  {\rm and} \  \ \
x = (x^{\mu})_{\mu=1,...,5} \in \R^5 \; ,
\een
and their configuration space is $X = \mathbb{R}^4 \times \mathbb{R}^5 $. 
Let $p_j$, $j=1,...,4$, and $p^{\mu}$, $\mu=1,...,5$, be the associated 
canonical momenta obeying,
\ben
[q_j, p_k ] = i \delta_{jk} \; , \quad [ x^{\mu} , p^{\nu} ] = 
i \delta^{\mu \nu} \; .
\een
The
fermionic degrees of freedom are described by the real Clifford
generators
\ben
\lam_{a} \; , a = 1, ..., 8 \ \ \ {\rm and } \ \ \ \psi_{a} \; , a
= 1,...,8 \; ,
\een
i.e., $\lam_a^{\dagger} = \lam_a$, $\psi_a^{\dagger} = \psi_a$, and
\begin{eqnarray*}
\{ \lam_a , \lam_b \} = \delta_{a b} \; ,  \quad  \{ \psi_a ,
\psi_b \} = \delta_{a b} \; ,  \quad
& \{ \lam_a, \psi_b \} = 0 \; .
\end{eqnarray*}
(Here and below $\{ \, \cdot \,  , \, \cdot \, \}$ stands for the anticommutator.)
By $\mathcal{F}$ we denote the irreducible representation space of this
Clifford algebra. The dimension of $\mathcal{F}$ is $2^8$. 
We introduce as a  preliminary Hilbert space
\ben
\mathcal{H}_0 = L^2( X ; \mathcal{F} ) = L^2 ( X ) \otimes
\mathcal{F} \; .
\een
As given in Appendix A,
we choose an explicit real irreducible representation
\ben
\g^{\mu} = (\g^{\mu}_{ab})_{a,b=1,...,8} \;  , \qquad \mu = 1, ..., 5 \; , 
\een
of the gamma matrices in 5 dimensions, i.e., 
\ben
\{ \g^{\mu} , \g^{\nu} \} =
2 \delta^{\mu \nu} \; .
\een
Furthermore we consider the real 
$8 \times 8$ matrices
\ben
s^i = (s^i_{ab})_{a,b=1,...,8} \; , \qquad i = 1, ..., 4 \; ,
\een
as they are defined in Appendix A. We note that $s^1 = \one_{8
  \times 8}$ and $(s^l)^T = - s^l$ for 
$l=2,3,4$ and that each $s^i$ commutes with
the $\g$--matrices.
We define
\ben
D_{ab} = \frac{1}{2}(q^R s^2 \overline{q}^R)_{ab} \; ,
\een
with 
\begin{eqnarray*}
q^R & = & s^1q_1 + s^2q_2 + s^3 q_3 + s^4 q_4 \; , \\
\overline{q}^R & = & s^1 q_1 - s^2 q_2 - s^3 q_3 - s^4 q_4 \; .
\end{eqnarray*}
We will use the convention of summing over repeated indices.
The supercharges are given by
\ben
Q_{a} =  (s^j \psi )_a p_j +   (\g^\mu \lam)_a p^{\mu} + D_{ab} \lam_b  + 
 (\g^{\mu} s^j s^2 \psi )_a x^{\mu} q_j      
\; , \qquad a = 1, ..., 8 \; .
\een
Note, for any $8 \times 8$ matrix $A$
we set $(A\psi)_a = A_{ab} \psi_b$, $\psi A \psi = \psi_a A_{ab} \psi_b$, 
and likewise for expressions containing $\lambda_a$.
The Hilbert space $\HH_0$ carries a unitary representation of ${\rm
  U}(1)$, called the gauge transformation,  
defined
by the generator
\ben
J = W_{12} + W_{34} - \frac{i}{2} \psi s^2 \psi \; ,
\een
where $W_{ij} = q_i p_j - q_j p_i$. 
We set $|x|:= (x^{\mu} x^{\mu})^{1/2}$ and $|q|:= (q_i q_i)^{1/2}$.
The full model satisfies
\begin{eqnarray}   \label{eq:mo:su}
\{ Q_a , Q_b \} =  \delta_{ab} H + 2 \g^{\mu}_{ab} x^{\mu} J \; ,
\end{eqnarray}
with
\begin{eqnarray*}
H & = &  p^{\mu} p^{\mu} + p_i p_i + |x|^2
|q|^2 + \frac{1}{4}|q|^4 - i x^{\mu} \psi \g^{\mu} s^2 \psi + i 2
q_{j} \lam s^j s^2 \psi \\
& = & -  \Delta + V + H_F \; ,
\end{eqnarray*}
where we have defined 
\ben
V =  |x|^2
|q|^2 + \frac{1}{4}|q|^4 \; , \  \mathrm{and} \ \  
H_F =  - i x^{\mu} \psi \g^{\mu} s^2 \psi + i 2
q_{j} \lam s^j s^2 \psi \; . 
\een
The Hilbert space of the model $\HH$ is the
${\rm U}(1)$--invariant subspace of $\mathcal{H}_0$, i.e.,
\ben
\HH = \{ \Psi \in \HH_0 \ | \ J \Psi = 0 \ \} \; .
\een
Note that the supercharges $Q_a$ are ${\rm U}(1)$ invariant and  
that on $\HH$ the superalgebra (\ref{eq:mo:su}) closes, i.e.,
\ben
\{ Q_a , Q_b \} \vert_{\HH} = \delta_{ab}  H \vert_{\HH} \; .
\een
The Hilbert space $\HH_0$ carries a natural representation of $\mathit{Spin}(5)$
defined by the infinitesimal generators
\ben
T^{\mu \nu} = x^{\mu} p^{\nu} - x^{\nu} p^{\mu} - 
\frac{i}{4} \gamma^{\mu \nu}_{ab} ( \lambda_a \lambda_b + \psi_a \psi_b) \; , 
\ \ \mu,  \, \nu = 1 , ..., 5 \ ,
\een
with $\gamma^{\mu \nu} = \frac{1}{2} [ \gamma^{\mu} , \gamma^{\nu} ]$.
Under this representation the supercharges $Q_a$ transform as spinors and the 
Hamiltonian $H$ is invariant. The action of $\mathit{Spin}(5)$
commutes with the gauge transformation, and thus leaves the
Hilbert space $\HH$ invariant.  

We introduce the fermionic number operator  $(-1)^F := 2^{8} \lam_1
\lam_2 ... \lam_8 \psi_1 \psi_2 ... \psi_8$, which anti-commutes
with $Q_a$ and commutes with $H$,
and  decompose the Hilbert space by means of  $(-1)^F$ as
\ben
\HH_{\pm} := \{ \Psi \in \HH \ | \ (-1)^F \Psi = \pm \Psi \ \} \; ,
\een
i.e., into bosonic (+) and fermionic (--) sectors.

We note that the operators $Q_a$ and $H$ are essentially self adjoint on
$C_0^{\infty}(X;\FF)$. Furthermore their restriction to $\HH$ is
essentially self adjoint on the space of 
${\rm U}(1)$--invariant functions in $C_0^{\infty}(X;\FF)$. 

\section{Result and outline of the proof}

The main Theorem is the following:

\begin{theorem} \label{secres:thm1}
There exists a state $\Psi \in \HH$ with $H \Psi = 0$.
\end{theorem} 
To prove this theorem, we use the deformation method 
introduced in \cite{porroz:bou}.
We consider the ``complex'' supercharges $D$ and $D^{\dagger}$,
\ben 
D  =  \frac{1}{\sqrt{2}} ( Q_1 + \i Q_2 ) \; , \qquad 
D^{\dagger}  =  \frac{1}{\sqrt{2}}(Q_1 - \i Q_2 ) \; .
\een
On  $\mathcal{H}$, $D^2 = {D^{\dagger}}^2 = 0$ and    
\begin{equation}  \label{eq:hequalsdd}
H = \{ D, D^{\dagger} \}  \; .
\end{equation}
We define the ${\rm U}(1)$--invariant function $w_k$ on $X$, by
\ben
w_k = k \cdot x^1 \; , \qquad {\rm for} \ \ k \geq 0 \; .
\een
We introduce the deformed supercharges
\ben
D_k = {e}^{-w_k} D { e}^{w_k} \; , \qquad 
D^{\dagger}_k = { e}^{w_k} D^{\dagger} { e}^{-w_k} \; .
\een
We
have
\begin{eqnarray*}
D_k = D - k \frac{\i}{\sqrt{2}}( ( \g^1 \lam)_1 + \i (\g^1 \lam )_2 )
\; , \quad
D^{\dagger}_k = D^{\dagger} + k \frac{\i}{\sqrt{2}}
( ( \g^1 \lam)_1 - \i (\g^1 \lam )_2 )
\; .
\end{eqnarray*}
As a little calculation shows, we have on $\HH$
\begin{eqnarray*}
H_k  =   \{ D_k , D^{\dagger}_k \}  \; , \quad 
\mathrm{with} \quad 
H_k := H +  k^2 +  k (q_3^2 + q_4^2 ) - 
 k (q_1^2 + q_2^2 )  \; .
\end{eqnarray*}
We point out that the deformed Hamiltonian is $\mathit{Spin}(5)$ invariant,
despite that the function $w_k = k \cdot x^1$ is not. 

\vspace{0.3cm}

The claim of Theorem \ref{secres:thm1} is an immediate consequence of
the following three propositions.

\begin{proposition} \label{secres:lem1}
If for some $k$ there exists a state $\Psi_k \in \HH$ 
with $H_k \Psi_k = 0$ such that
$\e^{\pm w_k} \Psi_k \in \HH$, then $H \Psi = 0$ for some  state 
$\Psi \in \HH$.
\end{proposition}

\vspace{0.3cm}
\noindent {\bf Remark.} Proposition \ref{secres:lem1} holds
for more general supersymmetric quantum mechanical systems 
and deformations, c.p.  \cite{porroz:bou}.

\vspace{0.3cm}
\noindent
The proof of Proposition \ref{secres:lem1}, which is presented in 
Subsection 
\ref{section4.1}, makes use of the Hodge decomposition and a cohomology 
argument.

\begin{proposition}  \label{secres:lem2}
For $k$ large enough, there exists a unique state $\Psi
\in \HH$ with $H_k \Psi = 0$. 
\end{proposition}

\vspace{0.3cm}
\noindent {\bf Remark.} Proposition \ref{secres:lem2} implies 
that $H_k$ admits a zero energy ground state for all $k > 0$. 
This follows from the stability of the Fredholm index 
of the continuous family of Fredholm operators,
$
(0,\infty) \ni  k \mapsto 
A_k := 2^{-1/2} (D_k +
D_k^{\dagger} )|_{\HH_{-}} \;  : \HH_- \to \HH_+  \; ,
$
where the topology is given by the graph norm with respect to $A_0$,
see for example  \cite{gil:inv}. However, we will not
use this fact to prove Theorem \ref{secres:thm1}.

\vspace{0.3cm}
\noindent
To prove Proposition  \ref{secres:lem2}, which is done in Subsection 
\ref{section4.2}, we first observe that the set of points, in which the scalar 
potential of the deformed Hamiltonian, i.e.,
\ben
V_k = V + k^2  - k (q_1^2 + q_2^2) + k(q_3^2 + q_4^2) \; ,
\een
vanishes, is a circle in configuration space $X$ 
(see e.g. (\ref{jps})). Its radius is proportional
to $k^{1/2}$. 
The circle is an orbit of the $U(1)$ action on $X$. In the direction 
orthogonal to the circle the Hessian of $V_k$ is non degenerate. 
Note that up to gauge transformations the scalar potential vanishes
exactly in one point.
Moreover, 
at infinity
the potential 
$V_k$ is bounded below by $k^2$. 
Using semiclassical 
analysis of eigenvalues, as given for example
in \cite{cycfrokirsim:sch}, together 
with a gauge fixing procedure, we show that there exists only one low 
lying eigenvalue for $k \to \infty$. 
In particular, we have to consider the fact that $H_F$ is unbounded from
below. By supersymmetry this low lying eigenvalue 
must equal zero for large $k$.

\begin{proposition}  \label{secres:lem3}
For $k>0$, a state $\Psi \in \HH$ with $H_k \Psi =0$ 
satisfies $\e^{\pm w_k} \Psi \in \HH$.
\end{proposition}

For the  proof of Proposition  \ref{secres:lem3},  which is given in 
Subsection \ref{section4.3},
we need to show that $\Psi$ decays sufficiently
fast as $|x| \to \infty$. We write the Hamiltonian as a sum of a 
free Laplacian in the $x$-variables and an $x$-dependent operator,
which describes the dynamics in the transverse direction. We show that the 
latter is bounded below by
$k^2 - c |x|^{-2}$ for some constant $c$ and $|x|$ large. Using an Agmon
estimate we then conclude that
\ben
|x|^{-1} e^{k|x|} \Psi
\een
is square integrable at infinity. As will be shown,
this together with the fact that $\Psi$
is invariant under $\mathit{Spin}(5)$ yields $e^{\pm w_k} \Psi \in \HH$.

\vspace{0.3cm}
\noindent {\bf Remark.}
To be precise, the operators $D, D^{\dagger}, D_k, D^{\dagger}_k$ and
$H_k$ are
defined in $\HH_0$ and $\HH$ as the closure on
$C_0^{\infty}(X;\FF)$  and $C_0^{\infty}(X;\FF) \cap
\HH$, respectively. The domain of $D$ is the set of all $\Psi$ in $\HH_0$ 
and $\HH$ such that 
$D\Psi$ (defined in the sense of distributions) is again in $\HH_0$ 
and $\HH$, respectively
(and analogous for the domains of $D^{\dagger}, D_k, D^{\dagger}_k,
 H$,  and $H_k$).
Indeed, $D^{\dagger}$ 
(resp. $D^{\dagger}_k$) is the adjoint of $D$ (resp. $D_k$).

\section{Proofs}

\subsection{Proof of Proposition \ref{secres:lem1}}
\label{section4.1}

We shall first show the Hodge decomposition
\begin{equation} \label{secpro:hod}
\HH = \ker H \oplus \overline{\ran D} \oplus \overline{\ran
  D^{\dagger}} \; . 
\end{equation}
To show the  orthogonality, we note that
\begin{equation*} 
(D \Psi , D^{\dagger} \Phi ) = ( D^2 \Psi , \Phi ) = 0 \; , 
\end{equation*}
with  $\Psi \in \DD(D)$ and $\Phi \in \DD(D^{\dagger})$, and
$\Psi \in \ker H$ iff $D \Psi = 0$ and $D^{\dagger} \Psi = 0$,
by (\ref{eq:hequalsdd}).
To show the completeness, we note that for each $\Psi \in (\ker
H)^{\perp}$,
\begin{eqnarray*}
\Psi & = & \lim_{a \downarrow 0} P_{(a, \infty )} (H) \Psi \\
& = &  \lim_{a \downarrow 0} \frac{1}{2}( D D^{\dagger} + D^{\dagger} D)
\frac{1}{H} P_{(a, \infty)} (H) \Psi \\
& = &  \lim_{a \downarrow 0} \left( \frac{1}{2} D ( D^{\dagger}
  \frac{1}{H} P_{(a,\infty)} (H) \Psi ) + 
\frac{1}{2} D^{\dagger} ( D \frac{1}{H} P_{(a,\infty)} (H) \Psi )
\right) \\
& = &  \lim_{a \downarrow 0} \frac{1}{2} D ( D^{\dagger}
  \frac{1}{H} P_{(a,\infty)} (H) \Psi ) + \lim_{a \downarrow 0}
\frac{1}{2} D^{\dagger} ( D \frac{1}{H} P_{(a,\infty)} (H) \Psi ) \; . 
\end{eqnarray*}
By $P_{\Omega}(H)$ we denoted the projection valued measure of
$H$, and the last equality follows since the two
terms belong to different orthogonal subspaces.  
Hence we have shown (\ref{secpro:hod}).

The equation $H_k \Psi_k = 0$ implies $D_k \Psi_k = 0$ and $D^{\dagger}_k
\Psi_k = 0$, and further $D e^{w_k} \Psi_k = 0$ and $D^{\dagger}
e^{- w_k} \Psi_k = 0$.
Assume $\ker H = \{ 0 \}$. Then 
\ben
e^{w_k} \Psi_k \in \ker D = {\ran D^{\dagger}}^{\perp} = \overline{\ran D}
\een
by the Hodge decomposition. 
It follows that  $\e^{w_k} \Psi_k = \lim_{n \to \infty} D
\Phi_n$ for some $\Phi_n$, but then
\begin{eqnarray*}
(\Psi_k, \Psi_k)   =  ( e^{w_k} \Psi_k , e^{- w_k } \Psi_k )  
 =  \lim_{n \to \infty} ( D \Phi_n , e^{- w_k} \Psi_k ) 
 =   \lim_{n \to \infty} ( \Phi_n , D^{\dagger} e^{- w_k} \Psi_k )  
 =  0 \; .
\end{eqnarray*}
This is a contradiction, and hence $\ker H \neq \{0\}$.  \qed

\subsection{Proof of Proposition \ref{secres:lem2}}
\label{section4.2}

We shall first rescale the operators $H_k$, $D_k$ and $D_k^{\dagger}$. For
$\Psi \in \HH$ and $t > 0$, we define the unitary operator
\ben
( U(t) \Psi )(\xi) = t^{9/2} \Psi (t \, \xi ) \; ,
\een
where $\xi = ( q, x)$. Furthermore, we define
\begin{eqnarray*}
K_t  & := & t^{2/3} U(t^{1/3}) H_{t^{2/3}} U^{*}(t^{1/3})  \\
F_t  & := & t^{1/3} U(t^{1/3}) D_{t^{2/3}} U^*(t^{1/3})   \\
F_t^{\dagger} & := & t^{1/3} U(t^{1/3}) D^{\dagger}_{t^{2/3}}
U^*(t^{1/3})  \; . 
\end{eqnarray*}
It follows that on $\HH$
\ben
\{ F_t , F_t^{\dagger} \} =  K_t  \quad  , \quad F_t^2 = 0 \quad , \quad
{F^{\dagger}_t}^2 = 0 \; , 
\een
and
\ben
K_t = - \Delta + t^2 V_1 + t H_F \; ,
\een
where
\ben
V_1 =  |x|^2|q|^2 + \frac{1}{4} |q|^4 + 1 + 
(q_3^2 + q_4^2 ) - (q_1^2 + q_2^2 ) \; .
\een
Proposition \ref{secres:lem2} follows from 

\begin{lemma} \label{lem:laszlo}
Let $E_n(t)$ denote the $n$th eigenvalue of $K_t$ counting
multiplicity. Then
\begin{equation} \label{secpro:cla}
\lim_{t \to \infty} E_1(t)/t = 0 \;  \ \ \textit{and} \ \ 
\liminf_{t \to \infty} E_2(t)/t \geq r > 0 \; .
\end{equation}
\end{lemma}
By supersymmetry, each non-zero eigenvalue of $K_t$ must be two fold
degenerate, i.e., occur as the eigenvalue of a pair consisting of a
bosonic and a fermionic eigenvector (see Theorem
6.3., \cite{cycfrokirsim:sch}). In view of
(\ref{secpro:cla}), for large $t$, two fold degeneracy of $E_1(t)$ is
not possible. Hence $E_1(t) = 0$. Moreover, 
this eigenvalue is nondegenerate.

\vspace{0.1cm}

\noindent
{\it Proof of Lemma  \ref{lem:laszlo}}. \ 
Writing the deformed potential $V_1$ as
\begin{equation}   \label{jps}
V_1 = |x|^2 |q|^2 +  \left( \sfrac{1}{2}(q_1^2 + q_2^2 ) - 1 \right)^2  +   
(q_3^2 + q_4^2)
\left(1 + \sfrac{1}{4}(q_3^2 + q_4^2) + \sfrac{1}{2}(q_1^2 + q_2^2) \right)
\end{equation}
we see that the set of points $\Gamma$ in which the potential
$V_1$ vanishes is given by
\begin{eqnarray*}
\Gamma & := & \{ (q,x) \in X \ | V_1(q,x)=0 \ \} \\
& = &
\{ (q,x) \in X \ | \ q_1^2 + q_2^2 = 2 , \ q_3=0, \ q_4=0, \ x = 0 \
\} \; .
\end{eqnarray*}
The set $\Gamma$ is a circle in the $(q_1 , \, q_2)$--plane about the
origin with radius
$\sqrt{2}$. The Hessian of $V_1$ at points lying in $\Gamma$ is
\begin{eqnarray*}
\left. (\mathrm{Hess} V_1)_{\a \b}  \right|_{\Gamma} = 
\left. \left( \frac{ \partial^2 V_1 }{ \partial \xi^{\alpha} \partial
      \xi^{\beta}} \right) \right|_{\Gamma} =
\left(
\begin{array}{ccc}
\left( 2 {q_r q_s} \right) & 0 & 0 \\
0 & 4 \one_{2 \times 2} & 0 \\
0 & 0 & 4 \one_{5 \times 5}   
\end{array}
\right) \; , \qquad \a,\b = 1,...,9 \; ,
\end{eqnarray*}
with $(\xi^1, ..., \xi^9 ): = ( q_1,..., q_4, x^1, ..., x^5)$ and $r,s
= 1,2$. At a point $p \in \Gamma$, the tangent to $\Gamma$ is
the only degenerate direction of the Hessian.

To show that there exists 
only one low lying eigenvalue, we will fix the 
${\rm U}(1)$ gauge.  
For $\omega \in L^1(X)$ with $ (W_{12} + W_{34}) \omega = 0$, we may integrate out
the coordinate $q_1$ as follows. We introduce the coordinates
\begin{eqnarray*}
&&\Phi : [0,2\pi] \times [0, \infty) \times \R^2 \longrightarrow \R^4   \\
&&\left( \begin{array}{c} \a \\  \rho \\ v_3 \\ v_4 \end{array} \right) 
\longmapsto 
\left( \begin{array}{c}
q_1 \\ q_2 \\ q_3 \\ q_4 \end{array} \right)  =
\left( \begin{array}{cc}
\left( \begin{array}{cc}
\cos \a & - \sin \a \\
\sin \a & \cos \a 
\end{array} \right)  & 0 \\
0 & 
\left( \begin{array}{cc}
\cos \a & - \sin \a \\
\sin \a & \cos \a 
\end{array} \right) \end{array} \right) 
\left( \begin{array}{c} 0 \\ \rho \\ v_3 \\v_4 \end{array}
\right) \; ,
\end{eqnarray*}
with $\alpha = {\rm arctan} (q_2/q_1)$ and 
$\rho = (q_1^2 + q_2^2)^{1/2}$. The
metric determinant is $\sqrt{ \det D \Phi^T D \Phi } = | \det D \Phi
| = \rho$, and
\ben
\int_{\R^4 \times \R^5 } \d q_1 ... \d q_4  \d^5 x \omega (q,x) = 
2 \pi \int_{ (0, \infty ) \times \R^2 \times \R^5 }\d \rho \d v_3 \d v_4 
\d^5 x  \rho \omega ((0, \rho ,v_3,v_4), x ) \; .
\een
The integration on the right hand side is reduced to the gauge fixed
configuration space $\widehat{X} := \{ 0 \} \times (0,
\infty ) \times \R^2 \times \R^5 \subset X $. We introduce the 
Hilbert space 
\ben
\widehat{\HH} := L^2 ( \widehat{X} ; \mathcal{F} )
\een
w.r.t. the Lebesgue measure of $\widehat{X}$, and we denote its canonical
scalar product by $\langle \ \cdot \ , \ \cdot \ \rangle_{GF}$. We 
define the isometry
\begin{eqnarray} \label{secprof:hat}
\HH & \longrightarrow & \widehat{\HH}  \\
\Psi & \longmapsto & \widehat{\Psi} := \sqrt{{2\pi \rho}} \, \Psi
\vert_{\widehat{X}} \; . \nonumber
\end{eqnarray}
By $M = - \frac{i}{2} \psi s^2 \psi $ we denote the spin part of $J$.
From $\widehat{\Psi}$ we may recover $\Psi$ through
\beqn \label{eq:4.1}
\Psi(q,x) = \frac{1}{\sqrt{2 \pi \rho}} \e^{- \i \alpha M}
\widehat{\Psi}( 0, \rho , q_3 \cos \alpha  - q_4 \sin \alpha  ,
q_4 \cos \alpha  + q_3 \sin \alpha  , x) \; .
\eeqn
Under the isometry 
(\ref{secprof:hat}), the corresponding transformation for the operators $A
\in \mathcal{L}(\HH)$, i.e., $A \to \widehat{A}  \in
\mathcal{L}(\widehat{\HH})$, is characterized by 
\ben
\widehat{A} \widehat{\Psi} = \widehat{A \Psi} \; .
\een
For $f \in
C_0^{\infty}(X)$, one has 
\beqn \label{p:pol3:dfdq}
\left. \left( \der{q_1} f \right) \right|_{\widehat{X}} = \left. \left(
   - \frac{i}{q_2} 
  W_{12} f \right) \right|_{\widehat{X}} \; ,
\eeqn
where the function $f$ is restricted to $\widehat{X}$ only after the
derivatives are performed. Applying this result to the function
$\partial f / \partial q_1$, using the commutation relation $[ W_{12},
  \partial / \partial q_1 ] = i \partial / \partial q_2$ and again 
(\ref{p:pol3:dfdq}), one finds 
\ben
\left.  \left( \der{q_1} \der{q_1} f \right) \right|_{\widehat{X}} =
\left. \left( 
    \frac{1}{q_2} \der{q_2} - \frac{1}{q_2^2} W_{12}^2  \right) f
 \right|_{\widehat{X}} \; .
\een
We set  $L := J - W_{12}$. Then for $\Psi \in \HH$,  
\ben 
W_{12} \Psi = - L \Psi \; .
\een
Note that $\widehat{L} = v_3 (-i \partial/ \partial v_4) -  
v_4 (-i \partial/ \partial v_3)  - \frac{i}{2} \psi s^2 \psi$.
For $\Psi \in \HH \cap C_0^{\infty}(X;\FF)$, a
straightforward calculation yields 
\ben
{\left( - \der{q_1} \der{q_1} \Psi  \right)}^{ \hspace{-3mm}
  \widehat{\quad}}   =  
- \frac{1}{\rho} \der{\rho} \widehat{\Psi} + \frac{1}{2 \rho^2} \widehat{\Psi} +
\frac{1}{\rho^2} \widehat{L}^2\widehat{\Psi} \; 
\een
and 
\ben
\left( - \der{q_2} \der{q_2} \Psi \right)^{ \hspace{-3mm} \widehat{
    \quad }} = -  
  \der{\rho} \der{\rho} 
    \widehat{\Psi} 
 + \frac{1}{\rho}
\der{\rho} \widehat{\Psi} -\frac{3}{4} \frac{1}{\rho^2 }
\widehat{\Psi}    \; . 
\een 
As a result 
\begin{equation}   \label{secpro:red}
({- \Delta \Psi})^{ \hspace{-0.5mm} \widehat{ \ }} = \left( -
  \Delta_{\widehat{X}} +  \rho^{-2} \left( \widehat{L}^2 - 
\textrm{\footnotesize $\frac{1}{4}$} 
\right) \right) \widehat{\Psi} \; ,
\end{equation}
where $\Delta_{\widehat{X}}$ is the formal Laplacian on $\widehat{X}$.
We will use eq. (\ref{secpro:red}) only for functions
in $C_0^{\infty}(\widehat{X}; \FF)$. 

We use the following partition of unity. We define
\ben
j_{1,t}(\xi) = \chi_r ( t^{2/5}( ({q_1^2 + q_2^2})^{1/2} - \sqrt{2})) \cdot
\chi_a ( t^{2/5} (q_3,q_4,x) ) \; , \quad \xi=(q,x) \; ,
\een
where for $\a = r,a$, we have chosen rotation invariant functions
$\chi_{\a} \in C_0^{\infty}(\R^{n_{\a}})$ with $n_r = 1$, $n_a = 7$,
$0 \leq \chi_{\a} \leq 1$, $\chi_{\a}(x) = 1$ if $|x| \leq 1$ and
0 if $|x| \geq 2$. Let $R \geq 1$ be fixed as $t \to \infty$.
We choose $j_2 \in
C^{\infty}(X)$ with $j_2(\xi) = j_2 (|\xi|)$, $0 \leq j_2 \leq 1$,
$j_2(\xi)=1$ for $|\xi| \geq 2 R$ and $j_2(\xi)=0$ for $|\xi| <
R$. Furthermore we set
\ben
j_{0,t} := ({ 1 - j_{1,t}^2 - j_2^2 })^{1/2} \; .
\een
For technical matters we consider the embedding $\widehat{X} 
\hookrightarrow \widetilde{X} := \{ 0 \} \times \R^8$ and the coordinates
$(0,\eta^2, ... , \eta^9 ) \in \widetilde{X}$.
By $\eta_0$ we denote the intersection of $\widehat{X}$ with $\Gamma$, i.e., 
$\eta_0  = ( 0 , \sqrt 2 , 0 , ... , 0 )$.
We define
\ben
\ov{V}_1^0(\eta ) = 
\frac{1}{2} \sum_{\a, \b = 2}^{9}
( \mathrm{Hess} V_1)_{\a \b}(\eta_0) 
(\eta^{\a} - \eta_0^{\a})  (\eta^{\b} -
\eta_0^{\b})\; . 
\een
and introduce the following operator on $L^2(\ov{X}  ; \mathcal{F} )$
\ben
G_t = - \Delta_{\ov{X}} + t^2 \ov{V}_1^0 + t H_F(\eta_0) 
\; , 
\een
where $H_F(\eta_0) = - i 2 \sqrt{2} (\lam_1 \psi_1 + ... \lam_8 \psi_8)  
: \FF \to \FF$ denotes the evaluation of $H_F$ at $\eta_0$
and $- \Delta_{\widetilde{X}}$ the eight dimensional 
Laplacian on $\widetilde{X}$.
For $\chi \in L^2(\ov{X} ; \mathcal{F}) $, we define the unitary 
transformation
\ben
(T(t) \chi )(\eta) = t^{2} \chi ( t^{1/2} (\eta -
\eta_0 ) ) \; .
\een
Then
\begin{eqnarray*} 
\frac{1}{t} T(t)^* G_t  T(t) =  - \Delta_{\ov{X}} +
 \frac{1}{2} (\mathrm{Hess} V_1)_{\a \b} (\eta_0)    
\eta^{\a} \eta^{\b} + H_F(\eta_0)  \; .
\end{eqnarray*}
The eigenvalue problem for this operator  
can be solved 
easily.
It has
purely discrete  
spectrum and its ground state  ${\Phi}^0$    has zero
energy and is non degenerate:  the sum of the first two terms is a harmonic 
oscillator, which acts on 
$L^2(\widetilde{X})$ and has ground state energy $8 \sqrt{2}$,
and $ H_F(\eta_0)$ acts on $\FF$ and has a unique ground state 
with energy $-8 \sqrt{2}$, see Appendix B (i).

Define
\ben
\widehat{\Psi}_t := \widehat{j}_{1,t} T(t) \Phi^0 \in 
C_0^{\infty}(\widehat{X}; \FF)  \hookrightarrow L^2(\widetilde{X}; \FF) \; .
\een
We recall that the corresponding $U(1)$-invariant wave function in
$\Psi_t \in \HH$ is obtained using (\ref{eq:4.1}).
Calculating
the energy of this state, we find
\begin{eqnarray}
\langle \Psi_t, K_t \Psi_t \rangle & = & \big\langle \widehat{\Psi}_t ,
\widehat{K}_t
\widehat{\Psi}_t \big\rangle_{GF}   \nonumber \\
& = &  \big\langle \widehat{\Psi}_t ,
( - \Delta_{\widehat{X}} + {\rho}^{-2} ( \widehat{L}^2 -
    \sfrac{1}{4} ) + t^2 \widehat{V}_1 + t \widehat{H}_F ) 
\widehat{\Psi}_t \big\rangle_{GF}  \nonumber \\
& = & \big\langle \widehat{\Psi}_t,  G_t  \widehat{\Psi}_t
\big\rangle_{GF}
+ \big\langle \widehat{\Psi}_t , {\rho}^{-2} ( \widehat{L}^2 -
  \textrm{\footnotesize $\frac{1}{4}$} )
\widehat{\Psi}_t 
\big\rangle_{GF}  \label{proo:pol3:ptht} \\
&  &  + 
\big\langle \widehat{\Psi}_t , t^2 ( \widehat{V}_1 - \ov{V}_1^0 )
\widehat{\Psi}_t 
\big\rangle_{GF}  +
\big\langle \widehat{\Psi}_t , t( \widehat{H}_F  - \widehat{H}_F(\eta^0)) 
\widehat{\Psi}_t \big\rangle_{GF}  \nonumber \; .
\end{eqnarray}
For the first term in (\ref{proo:pol3:ptht}), we find for $t \to \infty$,
\begin{eqnarray*}
 \big\langle \widehat{\Psi}_t, G_t \widehat{\Psi}_t
\big\rangle_{GF}  & = &
\big\langle T(t) {\Phi}^0 , \w{j}_{1,t} G_t \w{j}_{1,t} T(t)
{\Phi}^0 \big\rangle_{GF}  \\
& = & \big\langle T(t) {\Phi}^0 , ( \sfrac{1}{2} {\w{j}_{1,t}}^{\,2}
  G_t + \sfrac{1}{2} G_t {\w{j}_{1,t}}^{\,2}  + |
  \nabla_{\ov{X}} 
  \w{j}_{1,t}|^2 ) T(t) {\Phi}^0 \big\rangle_{GF} \\
& = & O(t^{4/5}) \; ,
\end{eqnarray*}
where  we denoted the gradient on $\ov{X}$ by  $\nabla_{\ov{X}}$, and we
used that $G_t T(t) {\Phi}^0 = 0$ 
and $\|\nabla_{\ov{X}} j_{1,t} \|_{\infty}^2 =  O( t^{4/5})$.  By rotation 
invariance of $\Phi^0$ in the $v_3, v_4$ variables, the
second term in 
(\ref{proo:pol3:ptht}) is an order one term. The estimate 
\beqn 
| t^2 {\w{j}_{1,t}}^{\,2} (\widehat{V}_1 - \ov{V}_1^0)| \leq \const \cdot t^2
{\w{j}_{1,t}}^{\,2} |\eta - \eta_0|^3 \leq \const \cdot t^2 \cdot t^{-6/5}
\label{p:pol3:e1}
\eeqn
yields $\langle \w{\Psi}_t , t^2 (\w{V}_1 - \ov{V}_1^0 ) \w{\Psi}_t
\rangle_{GF} = O(t^{4/5})$. And a similar estimate,
\beqn
| t{\w{j}_{1,t}}^{\,2} (\widehat{H}_F(\eta_0) - \widehat{H}_F)| \leq \const 
\cdot t
{\w{j}_{1,t}}^{\,2}  | \eta - \eta_0 | \leq \const \cdot t \cdot
t^{-2/5} \; , \label{p:pol3:e2}
\eeqn
gives 
\ben
\big\langle \w{\Psi}_t, t ( \widehat{H}_F  - \widehat{H}_F(\eta_0)) \w{\Psi}_t
\big\rangle_{GF} = O(t^{3/5}) \; ,
\een
as $t \to \infty$.
Collecting terms, we find
\ben
\langle \Psi_t , K_t \Psi_t \rangle = O(t^{4/5}) \; ,
\een
which implies that $\lim_{t \to \infty}E_1(t)/t = 0$.
This shows the first part of  (\ref{secpro:cla}).

\vspace{0.3cm}

To prove the second part, i.e., 
\beqn \label{p:pol3:le2t}
\liminf_{t \to \infty} E_2(t) /t \geq r > 0 \; , 
\eeqn
it suffices to show that there exists an $r>0$ such that
\beqn  \label{p:pol3:htqr}
K_t \geq ( t \cdot r + o(t)) \one + R_t  \; ,
\eeqn 
where $R_t$ is a symmetric, rank one operator. To see this, suppose
(\ref{p:pol3:htqr}) holds. Let $\omega_{1,t}$ and $\omega_{2,t}$ be the
eigenvectors to the eigenvalues $E_1(t)$ and $E_2(t)$ of $K_t$,
respectively. There exists a $\omega_t \in \mathrm{Span}\{ \omega_{1,t},
\omega_{2,t} \}$ in the kernel of $R_t$. Hence
\ben
E_2(t) \| \omega_t \|^2 \geq \langle \omega_t , K_t \omega_t \rangle 
\geq ( t \cdot r  + o(t) ) \| \omega_t \|^2  \; 
\een
which implies  (\ref{p:pol3:le2t}).

To show (\ref{p:pol3:htqr}), we use the IMS localization formula
\begin{eqnarray} \label{p:pol3:imsh}
K_t & = & \sum_{a=0}^1 j_{a,t} K_t j_{a,t} + j_2 K_t j_2 
- \sum_{a=0}^1 | \nabla j_{a,t} |^2  - |\nabla j_2 |^2 \; .
\end{eqnarray}
Now, $\supp (j_{0,t}) \subset \{ \xi \in X | \dist(\xi , \Gamma ) \geq
t^{-2/5} \}$. We have $\|\nabla j_{a,t}\|_{\infty}^2 = O(t^{4/5})$ for
$a=0,1$, and  
$\|\nabla j_2\|_{\infty}^2 = O(1)$. We estimate
\begin{eqnarray}
j_{0,t}K_t j_{0,t} &\geq& t^2 j_{0,t} V_1 j_{0,t} + t j_{0,t} H_F j_{0,t}
\nonumber \\
&\geq& (t^2 t^{-4/5} c_V - t c_F ) j_{0,t}^2 \; , \quad \mathrm{for \
  some} \ c_V > 0, \, c_F > 0   \nonumber \\
&\geq& t \cdot r  j_{0,t}^2  \; , \quad \mathrm{for \ some} \ r>0 \; ,
 \label{p:pol3:**}
\end{eqnarray}
and for $t$ large.
By fixing the gauge, we have on $L^2(\widehat{X}; \FF)$
\begin{eqnarray*}
\w{j}_{1,t} \w{K}_t \w{j}_{1,t} &=& 
 \w{j}_{1,t} G_t \w{j}_{1,t}
+ \w{j}_{1,t} t^2(\w{V_1} - \ov{V_1}^0)  \w{j}_{1,t}  \\
&&  +   \w{j}_{1,t} t (\w{H}_F - \w{H}_F(\eta_0)) \w{j}_{1,t} 
 +  \w{j}_{1,t} {\rho}^{-2} ( -\sfrac{1}{4} + \widehat{L}^2) \w{j}_{1,t} \\
& \geq & \w{j}_{1,t} G_t \w{j}_{1,t} + O(t^{4/5}) \\
& \geq &  \w{j}_{1,t} \,  t  \, r \left(  1  - | T(t) {\Phi}^0
\rangle_{GF} \cdot {}_{GF}\langle  T(t) {\Phi}^0 | \right)  
 \w{j}_{1,t}
+ O(t^{4/5}) \; ,
\end{eqnarray*}
for some  $r > 0$, where we have used the positivity of $\widehat{L}^2$, the
estimates (\ref{p:pol3:e1},\ref{p:pol3:e2}), and the gap in the spectrum
of $G_t$. On $\HH$, this yields
\begin{equation} \label{p:pol3:**2}
j_{1,t} K_t j_{1,t} \geq t \cdot r  j_{1,t}^2 - 
t \cdot r  | {\Psi}_t \rangle \cdot
\langle {\Psi}_t | + O(t^{4/5})  \; .
\end{equation}
To estimate the term $j_2 K_t j_2$, we recall the explicit form of
$K_t$:
\begin{eqnarray*}
K_t &=& p_i p_i + p^{\mu} p^{\mu} + t^2 \left( |x|^2|q|^2 + 
\sfrac{1}{4}|q|^4 + 1 -
(q_1^2 + q_2^2 ) + (q_3^2 + q_4^2 ) \right) \\
& &  + t ( - i x^{\mu} \psi \g^{\mu}
s^2 \psi + 2 i q_j \lambda s^j s^2 \psi ) \; .
\end{eqnarray*}
We recall the notation $\xi = (q,x)$. 
Define a function $\theta \in C^{\infty}(X)$ with
$\theta(\xi) = \theta(|q|)$, $0 \leq \theta \leq 1$, 
$\theta(|q|)=1$ if $|q|>4$
and $\theta(|q|)=0$ if $|q|<3$. Define $\bar{\theta}:= \sqrt{1-\theta^2}$. 
Then
\ben
j_2 K_t j_2 = j_2 \theta K_t \theta j_2 + j_2 \bar{\theta} K_t
\bar{\theta} j_2 - j_2 (|\nabla \theta |^2 + |\nabla \bar{\theta}|^2 ) j_2
\; .
\een
The localization error gives order 1 contributions, i.e.,
$\|\nabla \theta \|_{\infty}^2 = O(1)$, $\|\nabla
\bar{\theta} \|_{\infty}^2 = O(1)$.
First we consider the case where $|q|$ is large and estimate
(see Appendix B (i) for the terms containing fermions)
\begin{eqnarray*}
&&p^{\mu} p^{\mu} \geq 0 \; , \quad  -i x^{\mu} \psi \g^{\mu}s^2 \psi
\geq - 4 |x| \; , \quad  2 i q_j \lambda s^j s^2 \psi \geq - 8 |q| \; , \\
&& p_i p_i + t^2 |x|^2|q|^2 + t ( - i x^{\mu}
\psi \g^{\mu}s^2 \psi ) \geq 
p_i p_i + t^2 |x|^2 |q|^2 - 4 t  |x| \geq 0 \; ,
\end{eqnarray*}
where the last inequality follows from the ground state energy
of the harmonic oscillator.
This yields
\begin{eqnarray*}
j_2 \theta K_t \theta j_2 &\geq& j_2 \theta ( t^2 (\sfrac{1}{4}
|q|^4 - |q|^2 + 1) - 8 t  |q| ) \theta j_2   \\
&\geq& t^2 \cdot c j_2^2 \theta^2 \; ,
\end{eqnarray*}
for some $c >0$ and $t$ sufficiently large. For points $\xi = (q,x) \in
\supp j_2$, if $|q| < 4$, then $|x|$ is large for sufficiently large $R$.
We have
\begin{eqnarray*}
j_2 \bar{\theta} K_t \bar{\theta} j_2 & \geq & j_2 \bar{\theta} (
p_i p_i + 
t^2 |x|^2 ( 1 - |x|^{-2} ) |q|^2 - 4 t |x| +
t^2 - 8 t |q| ) j_2 \bar{\theta}  \\
& \geq & j_2 \bar{\theta} ( t ( 4 |x| (1 - |x|^{-2})^{1/2} - 4|x| ) +
t^2 - 32 t ) j_2 \bar{\theta} \; \\
& \geq & t^2 \cdot c j_2^2 \bar{\theta}^2 \; ,
\end{eqnarray*}
for some $c > 0$ and $t$ sufficiently large.
Hence there exists an $r>0$ such that for large $t$,
\beqn \label{p:pol3:***}
j_2 K_t j_2 \geq t \cdot r j_2^2  \; .
\eeqn
Now,
inserting eqns. (\ref{p:pol3:**}--\ref{p:pol3:***}) into
(\ref{p:pol3:imsh}) yields  (\ref{p:pol3:htqr})
and 
therefore (\ref{p:pol3:le2t}). 
\qed

\subsection{Proof of Proposition \ref{secres:lem3}}
\label{section4.3}

We decompose the Hilbert space $\HH_0$ as a constant fiber direct integral 
\cite{reesim:ana}, 
with fiber $F:=L^2(\R^4; \FF)$, 
\begin{eqnarray*}
\HH_0 =\int^{\oplus}_{\R^5} F  dx \;   ,
\end{eqnarray*}
the isomorphism being $
\Psi   \mapsto  ( x \mapsto \Psi_x := \Psi ( \cdot , x ) )$.
The Hamiltonian has a direct integral decomposition,
\ben
H_k  = p^{\mu} p^{\mu} + \int^{\oplus}_{\R^5} H_{k,x} dx \; ,
\een
where the fibers $H_{k,x}$, acting on $F$, are given by 
\ben
H_{k,x} = H^0_x + \frac{1}{4}|q|^4 + 2 i  q_j \lam s^j s^2 \psi + k^2 -
k (q_1^2 + q_2^2) + k (q_3^2 + q_4^2) \; , 
\een
with
\ben
H^0_x := p_i p_i + |x|^2|q|^2 - i x^{\mu} \psi \g^{\mu} s^2 \psi \; .
\een

The scalar product, the norm and operator norm in $F$ will be denoted by
$(  \cdot  ,  \cdot  )_F$ and $\| \cdot \|_F$,
respectively. Let $P_x$ be the projection onto the eigenspace of
$H^0_x$ corresponding to its lowest eigenvalue, which is, in fact, zero. 
We set $P^{\perp}_x := 1 - P_x$, and we define the projection
\begin{equation}  \label{eq:proj1}
P = \int_{\R^5}^{\oplus} P_x dx \; ,
\end{equation}    
and its complement $P^{\perp} = 1 - P$.
As is shown in Appendix B (ii),
for $x \neq 0$, 
\ben
\ran P_x  = \{ \ \Xi_x \cdot \xi \ | \
\xi \in \FF \ \mathrm{with} \ (\overline{u} \psi) \xi = 0 , \ 
\forall \  u : 
-i \g^{\mu} x^{\mu} s^2 u = |x| u \} \; ,
\een
where 
\ben
 \Xi_x(q) := (|x|\pi)^{-1} \exp(-\frac{1}{2}|x||q|^2) \; .
\een

\begin{lemma} \label{l1}
There exists an $R > 0$ and a constant 
$c > 0$ depending on $k$,
such that for $|x| > R$ 
\ben
H_{k,x}  \geq  k^2 -
  c|x|^{-2 }  \; . 
\een
\end{lemma}

\begin{proof}
Since all elements in $\ran P_x$ are spherically symmetric in $q$ it 
immediately follows that
\begin{equation}   \label{eq:star}
P_x H_{k,x} P_x  \geq k^2 P_x \; .
\end{equation}
We estimate, c.p. Appendix B (i),
\ben
2i q_j \lam s^j s^2 \psi \geq - 8 |q| 
\; , \quad
\mathrm{and} \ \ 
- k (q_1^2 + q_2^2 ) \geq - |x|^{-1} k ( |x|^{-1} p_i p_i + |x| |q|^2 )
\; . 
\een
Hence 
\begin{eqnarray*}
H_{k,x} &\geq& |x| (1 - |x|^{-1} ) ( |x|^{-1} p_i p_i + |x| |q|^2 ) +
|x|^{-1} p_i p_i + |x||q|^2  \\
& & - i x^{\mu} \psi \g^{\mu} s^2 \psi - 8 |q| - 
|x|^{-1} k (|x|^{-1} p_i p_i + |x||q|^{2} ) + k^2 \\
&\geq& |x| (1- |x|^{-1} - k|x|^{-2} ) ( |x|^{-1}p_i p_i + |x| |q|^2) 
- i x^{\mu}\psi \g^{\mu} s^2 \psi - 16 |x|^{-1} + k^2 \; ,
\end{eqnarray*}
where we used $|x||q|^2 - 8|q| \geq - 16 |x|^{-1}$ in the last inequality. 
The range of $P_x^{\perp}$ is given by the closure of the set of linear 
combinations of 
states which are a product of an eigenstate of $p_ip_i + |x|^2|q|^2$ and an 
eigenstate of $-i x^{\mu} \psi \g^{\mu} s^2 \psi$, excluding states which 
are  a product of two ground states. Thus 
\begin{equation} \label{eq:starstar}
P_x^{\perp} H_{k,x} P_x^{\perp}  \geq (c^0 |x| + k^2 ) P_x^{\perp} \; ,
\end{equation}
for some $c^0 > 0$ and large $|x|$. 

Using that 
$\| |q|^{ \a} P_x \|_F \leq c_{\a } |x|^{- \a / 2 }$ from the 
Gaussian decay of states in $\ran P_x$, 
\begin{eqnarray*}
\left\| P_x^{\perp} H_{k,x} P_x  \right\|_F & = & \left\| P_x^{\perp} 
\left(  
|q|^4 /4  - k
(q_1^2 + q_2^2) + k (q_3^2 + q_4^2) + 2 i q_j \lam s^j s^2 \psi \right) 
P_x  
\right\|_F \\
& \leq &  c |x|^{-1/2} 
\end{eqnarray*}
for some $c > 0$ and large $|x|$. By the
self adjointness of $H_{k,x}$ also 
\begin{equation} \label{eq:starstarstar}
\left\| P_x H_{k,x} P_x^{\perp}  \right\|_F \leq  c |x|^{-1/2} \; .
\end{equation}
Let $u_x \in \DD(H_{k,x}) \subset F$. Then 
from (\ref{eq:star}), (\ref{eq:starstar}) and (\ref{eq:starstarstar})
\begin{eqnarray*}
( u_x , H_{k,x} u_x )_F &\geq&
k^2 \| u_x \|_F^2 + 
\left( \begin{array}{c} 
\|P_x u_x \|_F \\
\|P_x^{\perp} u_x \|_F 
\end{array} \right) 
 A_{k,x} 
\left( \begin{array}{c} 
\|P_x u_x \|_F \\
\|P_x^{\perp} u_x \|_F 
\end{array} \right) \\
&\geq& \left( k^2 + \inf_{\| \xi \| = 1} ( \xi , A_{k,x} \xi ) \right) \|
u_x \|_F^2 \; ,
\end{eqnarray*}
with 
\ben
A_{k,x} := 
\left( \begin{array}{cc} 
0 & - c |x|^{-1/2} \\
- c |x|^{-1/2} & c^0 |x| \end{array}
\right)  \; .
\een
We have
\ben
\inf_{\| \xi \| =1 }( \xi , A_{k,x} \xi ) 
\geq - c |x|^{-2} \; ,
\een
for some $c > 0$ and large $|x|$. Hence the Lemmma 
follows.
\end{proof}

Let $R \geq 1$ be as in Lemma \ref{l1}, and let $\eta : \R^5 \to \R$ be a
smooth function with $\eta(x) = \eta(|x|)$, $0 \leq \eta \leq 1$,
$\|\nabla \eta \|_{\infty} \leq 1$, $\eta(x) = 0$ for $|x| \leq R$ and $\eta(x) = 1$ 
for $|x| \geq 3 R$.

The deformed supercharge 
\begin{equation} \label{eq:dfsc}
Q_{1,k} = Q_1 + k(\g^1 \lam)_2 = \frac{1}{\sqrt{2}}(D_k + D_k^{\dagger})
\end{equation}
satisfies on $\HH$, $2 (Q_{1,k})^2 = \{ D_k , D_k^{\dagger} \} = H_k$. Hence
for $\Psi \in \HH$, $H_k \Psi = 0$ iff $Q_{1,k} \Psi = 0$.

\begin{lemma} \label{l2}
Let $\Psi \in \HH$ with $H_k \Psi = 0$. Then for any $\epsilon > 0$, 
$|x|^{-1/2 - \epsilon} e^{k|x| }\eta \Psi \in \HH$.
\end{lemma}

\begin{proof}
It is sufficient to show the claim for arbitrarily small $\epsilon$.
To prove the lemma, we use an Agmon estimate \cite{agmon:lec}.
Let $h: \R^5 \to [0,\infty)$ be a smooth function such that
the set 
\ben
K = \{ x \in \R^5 | \ k^2 - c |x|^{-2} - | \nabla h (x) |^2 < 0
\ \}
\een
is compact. Then, as we will show,
\begin{equation} \label{ee1}
\int_{\R^5} \eta^2 \| \Psi_x \|^2_F ( k^2 - c |x|^{-2} - | \nabla
h(x) |^2 ) e^{2h} dx \leq  M_0  \| \Psi \|^2  \; .
\end{equation}
where 
\ben
M_0 := \sup_{R \leq |x| \leq 3R} \left|(1 + 2 |\nabla h(x)
|)e^{2h(x)}\right|   < \infty \; .
\een
Define $h_{\a} := h (1 + \a h)^{-1}$. Then, by Lemma \ref{l1},
\begin{eqnarray}  \label{ee2}
( \eta e^{h_{\a}} \Psi , H_k \eta e^{h_{\a}} \Psi ) & \geq &
\int_{\R^5} ( \eta e^{h_{\a}} \Psi_x , H_{k,x} \eta e^{h_{\a}} \Psi_x
)_F dx \\
& \geq & \int_{\R^5} \eta^2 e^{2 h_{\a}} ( k^2 - c |x|^{-2} ) \|
\Psi_x \|_F^2 dx \; . \nonumber
\end{eqnarray}
We estimate 
\begin{eqnarray*}
\left( \eta e^{h_{\a}} \Psi , H_k \eta e^{h_{\a}} \Psi \right) &=& 
2 \left( Q_{1,k} \eta e^{h_{\a}} \Psi ,  Q_{1,k} \eta e^{h_{\a}} \Psi
\right)   \\ 
&=& 2 \left( [Q_{1,k} , \eta e^{h_{\a}} ]  \Psi ,  [Q_{1,k} , \eta
  e^{h_{\a}} ]  \Psi   \right) \\
&\leq& \left( | \nabla (\eta e^{h_{\a}} )|^2 \Psi , \Psi \right) \\
&\leq& \left( (|\nabla \eta |^2 + 2 (\nabla \eta ) (\nabla h_{\a}) \eta +
  | \nabla h_{\a}|^2 \eta^2 ) e^{2 h_{\a}} \Psi , \Psi \right) \; .
\end{eqnarray*}
Inserting this into inequality (\ref{ee2}), we obtain
\begin{eqnarray*}
I_{\a} &:=& \int_{\R^5} \eta^2 e^{2 h_{\a}} (k^2 - c |x|^{-2} - | \nabla
    h_{\a}|^2 ) \| \Psi_x \|_F^2 dx  \\
&\leq& \left( (| \nabla \eta|^2 + 2 | \nabla \eta | | \nabla h_{\a}|
  \eta ) e^{2 h_{\a}} \Psi , \Psi \right) \\
&\leq& M_0  \| \Psi \|^2  \;  .
\end{eqnarray*}
Using Fatou's Lemma on the set $K^c$ and dominated convergence on $K$
yields
\ben
\left( \int_K + \int_{K^c} \right) \eta^2 \| \Psi_x \|_F^2 ( k^2 - c
|x|^{-2} - | \nabla h|^2 ) e^{2h} dx \leq \liminf_{\a} I_{\a} 
\leq
M_0  \| \Psi \|^2 
\een
and hence (\ref{ee1}). 

We choose $h$ such that on the support of $\eta$, 
$h(x) = k |x| - \epsilon \log |x|$. 
Then
\ben
k^2 - c |x|^{-2} - | \nabla h (x)|^2 = 2 k \epsilon |x|^{-1} - (c +
\epsilon^2) |x|^{-2}  \geq k \epsilon |x|^{-1} \; ,
\een
for large $|x|$. Hence, by (\ref{ee1})
\ben
\int_{\R^5} dx \eta^2 \| \Psi_x \|_F^2 e^{2k|x|}  k \epsilon
|x|^{-2\epsilon -1 }  < \infty \; ,
\een
which proves the Lemma.
\end{proof}

\vspace{0.5cm}

\noindent
{\it Proof of Proposition \ref{secres:lem3}}.
We recall that the deformed Hamiltonian commutes with the 
action of $\mathit{Spin}(5)$. Let $k$ be sufficiently large
such that $\Psi$ is the unique zero energy state of 
$H_k$. Thus $\Psi$ belongs to a one dimensional representation of
$\mathit{Spin}(5)$,
and therefore it is $\mathit{Spin}(5)$ invariant.
Let $R(S)$ denote the image of $S$ under the canonical
projection $\mathit{Spin}(5) \to SO(5)$. By $\RR(S)$ we  
denote the spin part of the  $\mathit{Spin}(5)$ action, i.e.,
the representation generated by  $- \frac{i}{4} \gamma^{\mu \nu}_{ab}  
( \lambda_a \lambda_b + \psi_a \psi_b)$. Then  
\ben 
\Psi(q, R(S)x) =  \RR(S) \Psi(q,x) \; , \quad \forall \ S \in \mathit{Spin}(5) \; . 
\een
This implies that for $x, x' \in \mathbb{R}^5$ with $|x|=|x'|$,
\ben
| \Psi_x |_{F} = | \Psi_{x'} |_{F}  \; .
\een
We set $\omega = x/|x|$. Let $d\Omega$ denote the surface measure of
the unit sphere. Then
\begin{eqnarray*} 
\int_{S^{4}}e^{- 2 k |x| \pm 2 k x^1} \,  d \Omega(\omega)
&=&
\mathrm{vol}(S^{3}) \int_{0}^{\pi} e^{- 2 k |x| ( 1 \mp \cos \theta
  )} \sin^{3} \theta d \theta \\ 
&\leq& \mathrm{vol}(S^{3}) \int_{-1}^1 e^{- 2 k |x| ( 1 - \cos \theta
  )}
 2 (1-\cos \theta) \,  d \cos \theta \\
&\leq& \mathrm{const} \ {|x|^{- 2}} \; . 
\end{eqnarray*}   
For the ground 
state $\Psi \in \HH$ of $H_k$, we have
\begin{eqnarray*}
\int e^{\pm 2 k x^1} |\Psi(q,x) |^2 \, dq dx &&=
\int e^{\pm 2 k x^1} |\Psi_x |_F^2 \,  dx \\
&&=  \int ( 1- \eta) e^{\pm 2 k x^1} |\Psi_x |_F^2 \,  dx +
 \int \eta e^{\pm 2 k x^1} |\Psi_x |_F^2 \,  dx  \\
&&\leq \mathrm{const}  +  
 \int \eta e^{\pm 2 k x^1} e^{- 2 k |x|} e^{ 2 k |x|} |\Psi_x |_F^2 \,  dx 
\\
&&\leq \mathrm{const} +  \mathrm{const} \int \eta |x|^{-2} e^{ 2 k |x|} 
|\Psi_x |_F^2 \,  dx \;  \\
&& < \infty  \ \ ,
\end{eqnarray*}
where in the last step we have used Lemma  \ref{l2}.
\qed

\vspace{0.5cm}

\noindent

\section*{Appendix A}

In this appendix we mainly follow \cite{setste:d0d}.
We consider the quaternions with generators $1,I,J,K$ satisfying the
relations
\ben
I^2 = -1 \; , \ \ J^2 = -1 \; , \ \ K^2 = -1 \; , \ \ I J K = -1 \; .
\een
A quaternion can be expanded as
\ben
q = q^1 1 + q^2 I + q^3 J + q^4 K \; .
\een
The conjugate is given by
\ben
\overline{q} = q^1 1 - q^2 I - q^3 J - q^4 K \; .
\een
We note that $q \overline{q} = \overline{q} q = |q|^2$.
By $1^R, I^R, J^R, K^R$ we denote the matrix representation, with
respect to the basis $(1,I,J,K)$, of the right multiplication with
$1,I,J,K$, respectively. We have
\begin{eqnarray*}
I^R = 
\left( 
\begin{array}{cccc} 
0 & -1 & 0 & 0 \\
1 & 0 & 0 & 0 \\
0 & 0 & 0 & 1 \\
0 & 0 & -1 & 0 
\end{array}
\right) 
\; ,  \quad 
J^R =  \left(
\begin{array}{cccc} 
0 & 0 & -1 & 0 \\
0 & 0 & 0 & -1 \\
1 & 0 & 0 & 0 \\
0 & 1 & 0 & 0 
\end{array}
\right) 
\; , \quad  
K^R =   \left(
\begin{array}{cccc} 
0 & 0 & 0 & -1 \\
0 & 0 & 1 & 0 \\
0 & -1 & 0 & 0 \\
1 & 0 & 0 & 0 
\end{array}
\right)  \; .
\end{eqnarray*}
Note that $(A B)^R = B^R A^R$ with $A,B \in \{
1, I, J, K \}$. 
We define the matrices
\begin{eqnarray*}
s^1 = 
\left(
\begin{array}{cc}
1^R & 0 \\
0  & 1^R 
\end{array}
\right) \; , \; 
s^2 = 
\left(
\begin{array}{cc}
I^R & 0 \\
0  & I^R 
\end{array}  
\right) \; , \; 
s^3 = 
\left(
\begin{array}{cc}
J^R & 0 \\
0  & J^R 
\end{array}
\right) \; , \;
s^4 =
\left(
\begin{array}{cc}
K^R & 0 \\
0  & K^R 
\end{array}
\right) \; . \ 
\end{eqnarray*}
We remark that $(s^l)^T = - s^l$ for $l=2,3,4$. We choose the gamma
matrices as
\begin{eqnarray*}
 \g^1 =
 \left(
\begin{array}{cc} 
\one_{4 \times 4} & 0 \\
0  & -\one_{4 \times 4} 
\end{array}
\right) \; , \;
\g^2 =
\left(
\begin{array}{cc}
0 & \one_{4 \times 4} \\
\one_{4 \times 4} & 0 
\end{array}
\right) \; , \; 
\g^3 = 
\left(
\begin{array}{cc}
0 & K^L \\
- K^L  & 0 
\end{array}
\right) \; , \;   
\end{eqnarray*}
\begin{eqnarray*}
 \g^4 = 
\left(
\begin{array}{cc}
0  & I^L \\
-I^L  & 0
\end{array}
\right) \; , \; 
\g^5 = 
\left(
\begin{array}{cc}
0  & J^L \\
-J^L  & 0
\end{array}
\right) \; , \; 
\end{eqnarray*}
with
\begin{eqnarray*}
I^L =  \left(
\begin{array}{cc}
-i \sigma^2  & 0 \\
0  & -i \sigma^2 
\end{array}
\right) \; , \;  
J^L =
 \left(
\begin{array}{cc}
0 & -\sigma^3 \\
\sigma^3  & 0
\end{array}
\right) \; , \ 
K^L =  
\left(
\begin{array}{cc}
0 & - \sigma^1 \\
\sigma^1 & 0 
\end{array}
\right) \; , \ 
\end{eqnarray*}
where $\sigma^i$, $i=1,2,3$, are the Pauli matrices
and the superscript $L$ indicates that the matrix corresponds to left
multiplication.
Using that left multiplication
commutes with right multiplication one sees that $[ \g^{\mu} , s^j ] =
0$. 

\vspace{0.5cm}

\noindent

\section*{Appendix B}

{\bf (i)} \,
Consider a real antisymmetric 
$16 \times 16$ matrix $S$ and the Clifford generators denoted as
$(\vartheta_1,...,\vartheta_8, \vartheta_9, ..., \vartheta_{16})=
(\psi_1, ... , \psi_8, \lam_1 , ... ,\lam_8)$. We will show that the map 
\ben 
i \sum_{a,b=1}^{16} \vartheta_a S_{ab} \vartheta_b : \FF \to \FF \;  
\een 
has a ground state $\xi \in \FF$, which is determined
by the condition that
\beqn  \label{appendixb:1}
\sum_{a=1}^{16} \overline{v}_a \vartheta_{a} \xi = 0 \; 
\eeqn 
for all eigenvectors $v$ of $iS$ with strictly positive
eigenvalue. The 
ground state energy is $-\frac{1}{2} \mathrm{tr} \sqrt{S^t S}$. If $S$ is 
invertible the ground state is unique. 

The matrix $iS$ is hermitian. Let $v$ be an eigenvector of $iS$ with
eigenvalue $\lambda$, then $\overline{v}$ is an eigenvector with
eigenvalue $-\lambda$. Hence we have the spectral decomposition
\ben
iS = \sum_{j=1}^8 \lambda_j ( P^{j +} - P^{j-} ) \; , \quad \lam_j \geq 0 \; ,
\een
where $P^{j \pm}$ are orthogonal projectors with 
${(P^{j \pm})}^t =  P^{j \mp}$. This yields
\begin{eqnarray*}
\sum_{a,b=1}^{16} \vartheta_a iS_{ab} \vartheta_b &=& 
\sum_{j=1}^8 \lam_j \sum_{a,b=1}^{16} 
\left( \vartheta_a  P^{j+}_{ab} \vartheta_b - \vartheta_a  P^{j-}_{ab} 
\vartheta_b 
\right) \\
&=& \sum_{j=1}^8 2 \lam_j \sum_{a,b=1}^{16}  
\vartheta_a  P^{j+}_{ab}  \vartheta_b  - \sum_{j=1}^8 \lam_j \; .
\end{eqnarray*}
Therefore, the ground state $\xi$ satisfies $(\ref{appendixb:1})$
and has energy
$- \sum_{j=1}^8 \lam_j  = - \frac{1}{2}\mathrm{tr} \sqrt{S^t S}$. If
$S$ is invertible then there are exactly $8$ linearly independent
eigenvectors with strictly positive eigenvalue. By the irreducibility of 
$\FF$, the condition
(\ref{appendixb:1}) then determines
the ground state uniquely.
 
\vspace{0.3cm}

\noindent
{\bf (ii)} \, 
Now, let us consider the special case $-i \psi x^{\mu} \gamma^{\mu} s^2 \psi$.
The vector $\xi \in \FF$ is a ground state of 
$-i \psi x^{\mu} \gamma^{\mu} s^2 \psi$ if and only if
\ben
(\overline{u} \psi ) \xi = 0 
\een
for all $u$ satisfying $-i \gamma^{\mu} x^{\mu}  s^2 u = |x| u$.
We define 
\ben
\WW_x := \{ \ \xi \in \FF \ | \ \xi \ \mathrm{is \ a \ ground \ state \
  of} \
- i \psi x^{\mu} \g^{\mu} s^2 \psi \ \} \; .
\een
The operators $\lam_a$ leave this space invariant and act irreducibly
on it. 
Thus $\dim \WW_x = 2^4$. 
The ground state of the harmonic oscillator
$p_i p_i + |x|^2 |q|^2$
is
\ben
\Xi_x(q) = (|x| \pi)^{-1} \exp ( - \sfrac{1}{2} |x|q^2) \; .
\een 
By $P_x$ we denote the projection onto the ground state of
\ben
H^0_x = p_i p_i + |x|^2 |q|^2  - i \psi x^{\mu} \g^{\mu} s^2 \psi \; .
\een
The harmonic oscillator part commutes with the fermionic part.
The ground state energy of $H^0_x$ is zero and 
\ben
\ran P_x = \{ \ \Xi_x \cdot \xi \ |  \ 
\xi \in \FF \ \mathrm{with} \ (\overline{u} \psi) \xi = 0 , \ 
\forall \  u : 
-i \g^{\mu} x^{\mu} s^2 u = |x| u \} \; .
\een

\vspace{0.5cm}

\noindent

\section*{Acknowledgement}
D.H. and L.E. want to 
thank the Mathematics Department
of the University of Copenhagen, at which this work was started.
J.P.S. wants to thank the Institute for Advanced Study, where part of this
work was done.
Moreover, D.H. wants to thank G.M. Graf, J. Hoppe, and
J. Fr\"ohlich for discussions.

\bigskip

\bibliography{mm}

\begin{thebibliography}{10}

\bibitem{porroz:bou}
M.~Porrati and A.~Rozenberg.
\newblock {B}ound {S}tates at {T}hreshold in {S}upersymmetric {Q}uantum
  {M}echanics.
\newblock {\em { Nucl. Phys. {\rm B}{\bf 515}}}, pages 184--202, 1998.
\newblock {\tt hep-th/9708119}.

\bibitem{c4a}
B.~de~Wit, J.~Hoppe, and H.~Nicolai.
\newblock On the quantum mechanics of supermembranes.
\newblock {\em { Nucl. Phys. {\rm B}{\bf 305}}}, pages 545--581, 1988.

\bibitem{c5a}
E.~Witten.
\newblock {B}ound {S}tates of {S}trings and {$p$}--{B}ranes.
\newblock {\em { Nucl. Phys. {\rm B}{\bf 460}}}, pages 335--350, 1996.
\newblock {\tt hep-th/9510135}.

\bibitem{c5b}
T.~Banks, W.~Fischler, S.H. Shenker, and L.~Susskind.
\newblock {M} {T}heory as a {M}atrix {M}odel: a {C}onjecture.
\newblock {\em Phys. Rev. D{\bf 55}}, pages 5112--5128, 1997.
\newblock {\tt hep-th/9610043}.

\bibitem{setste:aco}
S.~Sethi and M.~Stern.
\newblock {A} {C}omment on the {S}pectrum of {H}--{M}onopoles.
\newblock {\em {Phys.Lett.} {\rm B}{\bf 398}}, pages 47--51, 1997.
\newblock {\tt hep-th/9607145 }.

\bibitem{kacsmi:nor}
V.G. Kac and A.V. Smilga.
\newblock {N}ormalized {V}acuum {S}tates in {$N = 4$} {S}upersymmetric
  {Y}ang--{M}ills {Q}uantum {M}echanics with any {G}auge {G}roup.
\newblock {\em {Nucl. Phys. {\rm B}{\bf 571} }}, pages 515--554, 2000.
\newblock {\tt hep-th/9908096 }.

\bibitem{wit:sma}
E.~Witten.
\newblock {S}mall {I}nstantons in {S}tring {T}heory.
\newblock {\em { Nucl. Phys. {\rm B}{\bf 460}}}, page 541, 1996.
\newblock {\tt hep-th/9511030}.

\bibitem{berdou:fiv}
M.~Berkooz and M.~R. Douglas.
\newblock Five-branes in m(atrix) theory.
\newblock {\em Phys.Lett. B{\bf 395}}, pages 196--202, 1997.
\newblock {\tt hep-th/9610236}.

\bibitem{doukabpoushe:dbr}
M.~R. Douglas, D.~Kabat, P.~Pouliot, and S.~H. Shenker.
\newblock D-branes and short distances in string theory.
\newblock {\em Nucl.Phys. B{\bf 485}}, pages 85--127, 1997.
\newblock {\tt hep-th/9608024}.

\bibitem{pol:supII}
J.~Polchinski.
\newblock {\em String Theory}, volume Volume II.
\newblock Cambridge University Press, 1998.

\bibitem{setste:inv}
S.~Sethi and M.~Stern.
\newblock {I}nvariance {T}heorems for {S}upersymmetric {Y}ang-{M}ills
  {T}heories.
\newblock {\em {Adv. Theor. Math. Phys.} {\bf 4}}, pages 487--501, 2000.
\newblock {\tt hep-th/0001189}.

\bibitem{agmon:lec}
S.~Agmon.
\newblock {\em {L}ectures on {E}xponential {D}ecay of {S}olutions of
  {S}econd-{O}rder {E}lliptic {E}quations: {B}ounds on {E}igenfunctions of
  {N}-{B}ody {S}chr\"odinger {O}perators}.
\newblock Princeton {U}niversity {P}ress, 1982.

\bibitem{fghhy}
J.~Fr{\"o}hlich, G.M. Graf, D.~Hasler, J.~Hoppe, and S.-T. Yau.
\newblock Asymptotic form of zero energy wave functions in supersymmetric
  matrix models.
\newblock {\em {Nucl. Phys. {\rm B}{\bf 567}}}, pages 213--248, 2000.
\newblock { \tt hep-th/9904182}.

\bibitem{setste:d0d}
S.~Sethi and M.~Stern.
\newblock The {S}tructure of the {D0-D4} {B}ound {S}tate.
\newblock {\em {Nucl. Phys.} {\rm B}{\bf 578}}, pages 163--198, 2000.
\newblock {\tt hep-th/0002131}.

\bibitem{gil:inv}
P.~Gilkey.
\newblock {\em Invariance {T}heory, the {H}eat {E}quation, and the {A}tiyah-
  {S}inger {I}ndex {T}heorem}.
\newblock CRC Press, 2nd edition, 1994.

\bibitem{cycfrokirsim:sch}
H.L Cycon, R.G. Froese, W.~Kirsch, and B.~Simon.
\newblock {\em Schr{\"o}dinger Operators}.
\newblock {S}pringer--{V}erlag, 1986.

\bibitem{reesim:ana}
M.~Reed and B.~Simon.
\newblock {\em Methods of Modern Mathematical Physics, {IV} Analysis of
  Operators}.
\newblock Academic Press, New York, 1978.

\end{thebibliography}
\bibliographystyle{unsrt}



\end{document}